\documentclass{article}
\usepackage{spconf,amsmath,graphicx,color}
\usepackage{amsfonts}
\usepackage{amssymb}
\usepackage{url,subfigure,cite,array}
\usepackage{hyperref} 
\usepackage{color}
\usepackage{booktabs}
\usepackage{tabularx}
\usepackage{multirow}
\usepackage{adjustbox}

\title{PromptVC: Flexible Stylistic Voice Conversion in Latent Space Driven by Natural Language Prompts}
\name{
\begin{tabular}{c}
\it Jixun Yao$^1$, Yuguang Yang$^2$, Yi Lei$^1$, Ziqian Ning$^1$, Yanni Hu$^2$, Yu Pan$^2$, \\
\it Jingjing Yin$^2$, Hongbin Zhou$^2$, Heng Lu$^2$, Lei Xie$^{1,*}$\thanks{* Corresponding author},
\end{tabular}
}
\address{$^1$Audio, Speech and Language Processing Group (ASLP@NPU)\\School of Computer Science, Northwestern Polytechnical University, Xi’an, China\\
$^2$Ximalaya Inc, China}
\begin{document}
\ninept
\maketitle
\begin{abstract}
Style voice conversion aims to transform the style of source speech to a desired style according to real-world application demands. However, the current style voice conversion approach relies on pre-defined labels or reference speech to control the conversion process, which leads to limitations in style diversity or falls short in terms of the intuitive and interpretability of style representation.
In this study, we propose \textit{PromptVC}, a novel style voice conversion approach that employs a latent diffusion model to generate a style vector driven by natural language prompts. Specifically, the style vector is extracted by a style encoder during training, and then the latent diffusion model is trained independently to sample the style vector from noise, with this process being conditioned on natural language prompts.
To improve style expressiveness, we leverage HuBERT to extract discrete tokens and replace them with the K-Means center embedding to serve as the linguistic content, which minimizes residual style information. Additionally, we deduplicate the same discrete token and employ a differentiable duration predictor to re-predict the duration of each token, which can adapt the duration of the same linguistic content to different styles.
The subjective and objective evaluation results demonstrate the effectiveness of our proposed system.

\end{abstract}

\vspace{-10pt}
\section{Introduction}
\label{sec:intro}
\vspace{-10pt}
Voice conversion (VC) is a speech signal transformation technique that changes the original speaker's voice into the target voice while maintaining the linguistic content~\cite{sisman2020overview}. This technique has attracted broad interest in the audio and speech processing community due to its diverse applications in real-world scenarios. These applications encompass movie dubbing~\cite{bgmvc}, audiobook synthesis~\cite{audiobook}, and speaker anonymization for voice privacy protection~\cite{dis_anon}.
VC can convert distinct styles of the source speech into target styles (\textit{e.g.}, timbre, emotion, and speaking rhythm), resulting in a better listening experience in various application scenarios. However, all previous style voice conversion systems rely on auxiliary categorical labels or reference speech as conditions to control specific styles, which are time-consuming and not user-friendly. Therefore, it is a more user-friendly and flexible way if the conversion is achieved with a text description in natural language.


Mainstream studies on style voice conversion can be broadly categorized into two classes~\cite{emo_overview}: (1) categorical label-based VC~\cite{id1,id2} and (2) reference speech-based VC~\cite{ref1,ref2,ref3,ref4,ref5}. 
Categorical label-based methods involve a set of pre-defined auxiliary categorical labels to represent individual styles. These labels are then utilized to control the conversion process. On the other hand, reference speech-based methods eliminate the necessity for pre-defined labels. Instead, they employ a reference encoder, with Global Style Tokens (GST) being the most prominent example~\cite{gst}, to model an extensive range of expressive styles from reference speech. These methods convert the source style into the target style by using different reference speech during the inference stage.

Although these methods can effectively convert the original style in the source speech to the target style, categorical label-based methods have limitations in terms of the diversity of style expressiveness. The synthetic samples only contain some kind of "averaged" styles learned from the training data. While reference speech-based methods can be trained in an unsupervised manner and generate more expressiveness samples, the selection of reference speech requires careful consideration. Meanwhile, the style information contained in reference speech may not be intuitive or interpretable. 
On the other hand, if the intention is to simultaneously convert two style attributes, such as timbre and emotion, including an extra set of categorical labels or utilizing another reference speech becomes necessary~\cite{expressivevc}.
Although some methods in recent text-to-speech studies employ a text prompt to control speech synthesis~\cite{yang2023instructtts,promptstyle,guo2023prompttts}, this process is simpler when compared to VC. This is because VC needs to remove the original style information inherently in the source speech, a step not required in the case of text-to-speech.

\begin{figure*}[ht]
  \centering
  \includegraphics[width=15cm]{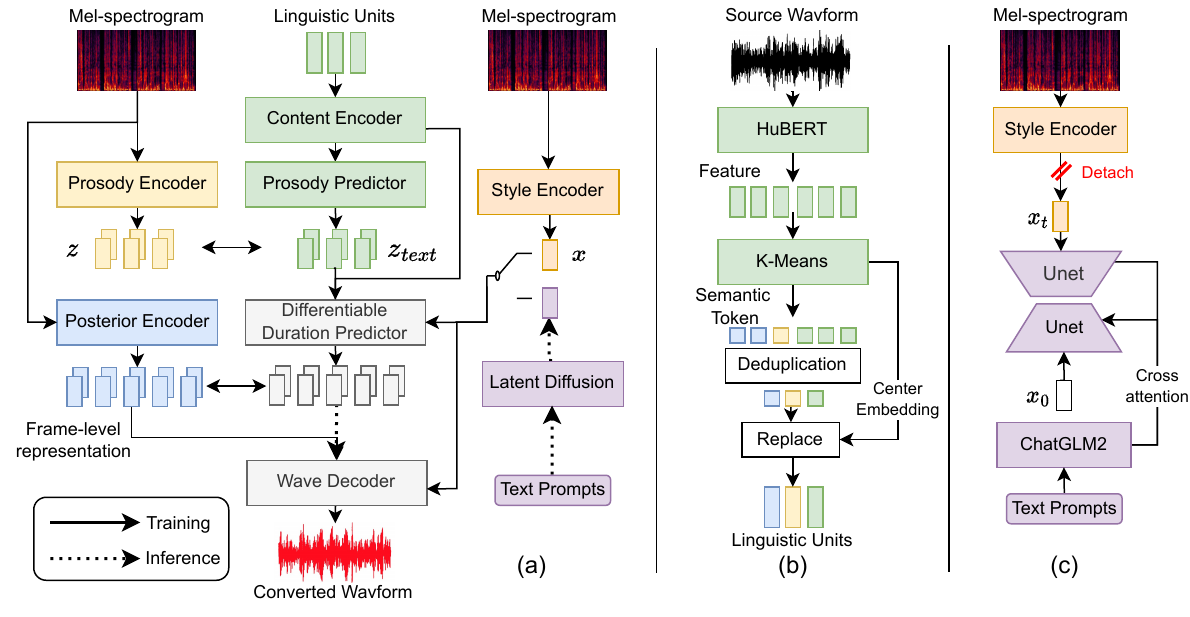}
  \caption{The details of our proposed approach. Subfigure (a) is the architecture of PromptVC. The solid line indicates the training stage while the dashed line represents the inference stage. Subfigures (b) and (c) illustrate the process of linguistic unit extraction and the training procedure of the latent diffusion model, respectively.}
    \vspace{-15pt}
  \label{fig:model}
\end{figure*}

In this study, to flexibly control the style conversion in an intuitive way, we propose \textit{PromptVC}, which employs a latent diffusion model to generate the style vector conditioned on natural language prompts. An end-to-end VC model is trained to reconstruct the target waveform controlled by the style encoder, while the latent diffusion model is trained to sample the output of the style encoder from noise. 
During the inference stage, the latent diffusion model is guided by natural language prompts to generate the target style vector and then the conversion model to reconstruct the speech conditioned on the generated style vector. 
To efficiently minimize residual style information in linguistic content, we leverage HuBERT~\cite{hubert} to extract discrete tokens from the source speech. These tokens are then deduplicated for the purpose of predicting the duration of each token through a differentiable duration predictor. Furthermore, we substitute the discrete tokens with K-Means center embeddings, which encompass more relative positional information about linguistic content and can help mitigate mispronunciation issues. For better prosody modeling and more natural-sounding speech, we introduce a prosody encoder to capture the phoneme-level prosody representation.
Experiments and ablation studies show the significant advantages of the proposed approach.


\section{Proposed Method}

\subsection{System Overview}
The overall architecture of PromptVC is illustrated in Figure~\ref{fig:model}(a), which can be expressed as a conditional variational autoencoder. In the training phase, the posterior encoder and prosody encoder convert the input mel-spectrogram to a sequence of frame-level latent variables and phoneme-level prosody representation, respectively. 
Subsequently, the content encoder models the high-level linguistic representations, which are utilized to predict the prosody representation by the prosody predictor. The differentiable duration predictor transforms the phoneme-level linguistic representation with prosody representation into the frame-level intermediate representations, constrained by the evidence lower bound (ELBO) of the marginal log-likelihood of the frame-level latent variable of target speech. The style encoder aims to extract a global vector that contains the style information of the input mel-spectrogram. This vector serves as a condition for the differentiable duration predictor and wave decoder. Finally, the wave decoder generates the waveforms from the frame-level latent variable. During inference, the target style vector is generated by the latent diffusion model using text prompts as a condition instead of extracting from the style encoder.
\subsection{Linguistic Units Extraction}
We introduce a novel linguistic content extraction method that leverages discrete tokens to disentangle the linguistic content from source speech and minimize residual style information, as illustrated in Figure~\ref{fig:model}(b). 
The discrete token, commonly referred to as a semantic token, is extracted from the source speech using a pre-trained HuBERT model with K-Means cluster model~\cite{hubert}. The HuBERT model is accessible on an open-source website~\footnote{https://github.com/TencentGameMate/chinese\_speech\_pretrain}, and we utilize the final layer of HuBERT to cluster the K-means model.
This frame-level token contains minimal content-independent information in contrast to the conventionally employed phonetic posteriorgrams (PPG)~\cite{ref5}. Furthermore, variations exist in the duration of the same linguistic content across different styles. However, traditional VC processes the speech into the same duration as the original speech. To adjust to diverse style expressions, we deduplicate the same semantic token for the purpose of predicting the duration of each token using the duration predictor. To avoid the mispronunciation problem, we substitute the deduplicated token with the corresponding center embedding, denoted as the linguistic unit, in the K-Means cluster model. The linguistic units contain more relative position information of linguistic content compared to discrete deduplicated tokens. The phoneme-level linguistic units extracted using this method minimize residual information from the original style and can be converted to diverse durations under distinct style conditions.
\subsection{Style Modeling}
The style encoder utilizes multi-head self-attention and temporal averaging to extract a global style representation from the reference mel-spectrogram $x_{\text{mel}}$. This representation serves as guidance to control the style of the output speech. The architecture of the style encoder is the same as~\cite{stylespeech}, and we also incorporate Style-Adaptive Layer Normalization (SALN) to align the gain and bias of the linguistic content based on the extracted style representation.

\begin{figure}[ht]
  \centering
  \includegraphics[width=6cm]{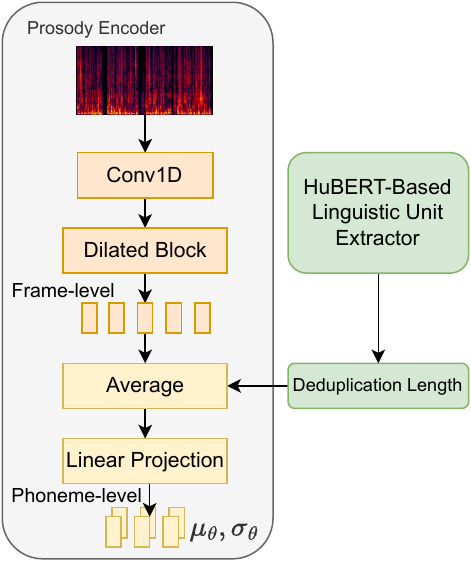}
  \caption{The model architecture of prosody encoder.}
  \label{fig:prosody}
  \vspace{-15pt}
\end{figure}

To generate more natural-sounding speech, we introduce an additional prosody encoder to capture the phoneme-level prosody information. As shown in Figure~\ref{fig:prosody}, the prosody encoder consists of 4 WaveNet residual dilated blocks~\cite{wavenet}, each comprising layers of dilated convolutions with a gated activation unit and skip connection. 
The output of the dilated block is a sequence of frame-level representation, which may cause linguistic content information leakages during training. To address this issue, we convert the frame-level representation to the phoneme-level prosody feature by utilizing the deduplication length as the phoneme-level duration.
Therefore, we can employ KL divergence to constrain the phoneme-level prosody representation extracted from both the prosody encoder and the prosody predictor in the following manner:
\begin{equation}
    \mathcal{L}_{\text{pro}}=\mathbb{E}_{q_\phi(z \mid x_{\text{mel}})} \left[\log q_\phi(z \mid x_{\text{mel}})-\log p_\theta(z \mid z_{\text{text}}) \right],
\end{equation}
where $z$ and $z_{\text{text}}$ represent the prosody representation obtained from the prosody encoder and prosody predictor, respectively. 
The prosody modeling enables our model can predict phoneme-level prosody during the inference stage and avoid the one-to-many mapping problem compared with directly predicting pitch and energy. 

We adopt the deduplication length as the duration, which is notably shorter than the conventional phoneme duration. However, employing conventional hard expansion techniques, like the FastSpeech duration predictor~\cite{fastspeech,fastspeech2}, may result in inaccurate duration predictions. Therefore, we utilize a differentiable duration predictor which includes a trainable upsampling layer. This layer employs the predicted duration to train a projection matrix, enabling the extension of the linguistic hidden sequence from phoneme level to frame level in a differentiable manner. The content encoder, posterior encoder, and wave decoder are the same architecture used in VITS~\cite{vits}. The training objectives remain consistent with the original VITS training goals while adding prosody loss $\mathcal{L}_{\text{pro}}$.


\subsection{Generative Latent Diffusion}
To achieve style voice conversion using natural language prompts, we employ a latent diffusion model~\cite{ldm} to generate the global style vector extracted from the style encoder, as shown in Figure~\ref{fig:model}(c). 
Conditioned on textual representation, the latent diffusion model divides the generation process into multiple conditional diffusion steps. Text-guided generation models need powerful semantic text understanding encoders to capture the meaning of natural language prompts.
We use the ChatGLM2-6B~\footnote{\url{https://github.com/THUDM/ChatGLM2-6B}} language model to generate textual representations from natural language prompts. This generated textual representation serves as the condition for the latent diffusion model, employing a cross-attention manner.

The input of the latent diffusion model, denoted as $\mathbf{x}_t$, is the noisy style vector in some steps, which is obtained by corrupting the original style vector $\mathbf{x}$ via a Gaussian diffusion process with a noise schedule parameterized by the standard deviation of the noise at time $t$.
The training loss of the latent diffusion model $\boldsymbol{\epsilon}_\theta$ is defined as the mean squared error in the noise space:
\begin{equation}
\mathcal{L}_{\text{diff}} =\left\|\boldsymbol{\epsilon}_\theta\left(\mathbf{x}_t, \mathbf{c}, t\right)-\boldsymbol{\epsilon}\right\|^2,
\end{equation}
where $\mathbf{c}$ is the textual representation and $\boldsymbol{\epsilon}$ is diffusion noise sampled from a standard normal distribution.

\section{Experiments}
\subsection{Datasets}
Since there is a lack of VC datasets that encompass expressive speech with comprehensive style prompts, we use an internally expressive multi-speaker Mandarin corpus to train the VC model. This corpus comprises 400 hours of speech data at a sampling rate of 24kHz. From this corpus, we specifically selected 50,000 utterances that encompass a range of characteristics, including six emotions, two distinct scenarios (news and novel), and six different timbres (three male and three female). Subsequently, we invite 13 professional annotators to provide annotations in the form of natural language prompts for these 50,000 utterances to train the diffusion model.

\subsection{Baseline and Evaluation Metrics}
To begin, we conduct an evaluation of our proposed PromptVC using reference speech to achieve style voice conversion and two models are established as the baseline. 
The first baseline system is StyleSpeech~\cite{stylespeech} and we replace the original phoneme input with linguistic units used in our approach. The second baseline system, called MixEmo~\cite{zhou2022mixed}, employs a ranking order methodology to represent the degree of relevance concerning emotion or style. 

We employ both subjective and objective metrics to compare our proposed PromptVC with the above baseline system across two aspects: speech quality and style similarity. For speech quality, we utilize various metrics including Mel-cepstral distortion (MCD)~\cite{mcd}, Short-Time Objective Intelligibility (STOI)~\cite{stoi}, Character Error Rate (CER), and Mean Opinion Score (MOS) test to evaluate speech quality. The CER is calculated using a pre-trained automatic speech recognition model provided by the WeNet toolkit~\footnote{\url{https://github.com/wenet-e2e/wenet}}. For style similarity, we employ two pitch-related metrics: Root Mean Squared Error (RMSE) and Pearson correlation (Corr)~\cite{cohen2009pearson}. These two metrics are widely applied to evaluate the performance of expressive VC. Since the sequences are not aligned, we perform Dynamic Time Warping to align the sequences prior to comparison.
Additionally, we conduct a Similarity Mean Opinion Score (SMOS) test to evaluate the perceived style similarity. In both the MOS and SMOS tests,  participants are presented with randomly selected speech samples and requested to score the speech quality or style similarity of each sample on a 5-point scale ('5' for excellent, '4' for good, '3' for fair, '2' for poor, and '1' for bad). Audio samples are available online~\footnote{\url{https://yaoxunji.github.io/prompt_vc/}}.

\begin{table*}[ht]
\centering
\tiny
\renewcommand\arraystretch{1.2}
\caption{Comparison results of speech quality and style similarity between our proposed PromptVC and baseline systems. MOS and SMOS are the subjective test results, presented with 95\% confidence intervals.}\label{tab:compare}
\resizebox{1.0\linewidth}{!}{
\begin{tabular}{l|cccc|ccc}
\hline
\multicolumn{1}{c|}{\multirow{2}{*}{Model}} & \multicolumn{4}{c|}{Speech Quality} & \multicolumn{3}{c}{Style Similarity} \\ \cline{2-8} 
\multicolumn{1}{c|}{}                       & MCD ($\downarrow$)  & STOI ($\uparrow$) & CER ($\downarrow$)  & MOS ($\uparrow$) & RMSE ($\downarrow$)     & Corr ($\uparrow$)    & SMOS ($\uparrow$)   \\ \hline
MixEmo ~\cite{zhou2022mixed}                                     & 7.93  & 0.56  & 9.16  & 3.08$\pm$0.07 & 16.63     & 0.78     & 3.29$\pm$0.07    \\
StyleSpeech ~\cite{stylespeech}                                & 6.24  & 0.71  & 7.43  & 3.69$\pm$0.05 & 11.24     & \textbf{0.88}     & 3.61$\pm$0.04    \\
PromptVC                                    & \textbf{5.91}  & \textbf{0.76}  & \textbf{6.91}  & \textbf{3.90$\pm$0.04} & \textbf{11.06}     & 0.86     & \textbf{3.86$\pm$0.06}    \\ \hline
\end{tabular}
}
\vspace{-15pt}
\end{table*}

\vspace{-10pt}
\subsection{Experimental Results}
\subsubsection{Style Voice Conversion by Reference Speech}
Since the baseline system utilizes reference speech to conduct style voice conversion, we first investigate the performance of PromptVC using reference speech. This serves as a preliminary step before moving on to the evaluation of using natural language prompts to convert style. Table~\ref{tab:compare} shows the evaluation results of speech quality and style similarity among the proposed PromptVC and the baseline systems. In terms of speech quality, PromptVC surpasses the compared systems by achieving lower MCD and CER scores, as well as higher STOI results. Furthermore, the MOS results indicate that PromptVC performs better in terms of perceptual quality. These findings highlight that PromptVC exhibits better speech quality performance in contrast to the baseline system.

On the other hand, PromptVC excels in RMSE and SMOS performance, particularly when compared to MixEmo, while PromptVC displays a pitch correlation result of 0.86, which is closely similar to the 0.88 observed in StyleSpeech. Furthermore, the SMOS results of PromptVC substantially outperform both baseline systems. This indicates that the converted speech generated by PromptVC exhibits greater similarity to the target speech in terms of style.

\vspace{-5pt}
\begin{figure}[ht]
  \centering
  \includegraphics[width=1.0\linewidth]{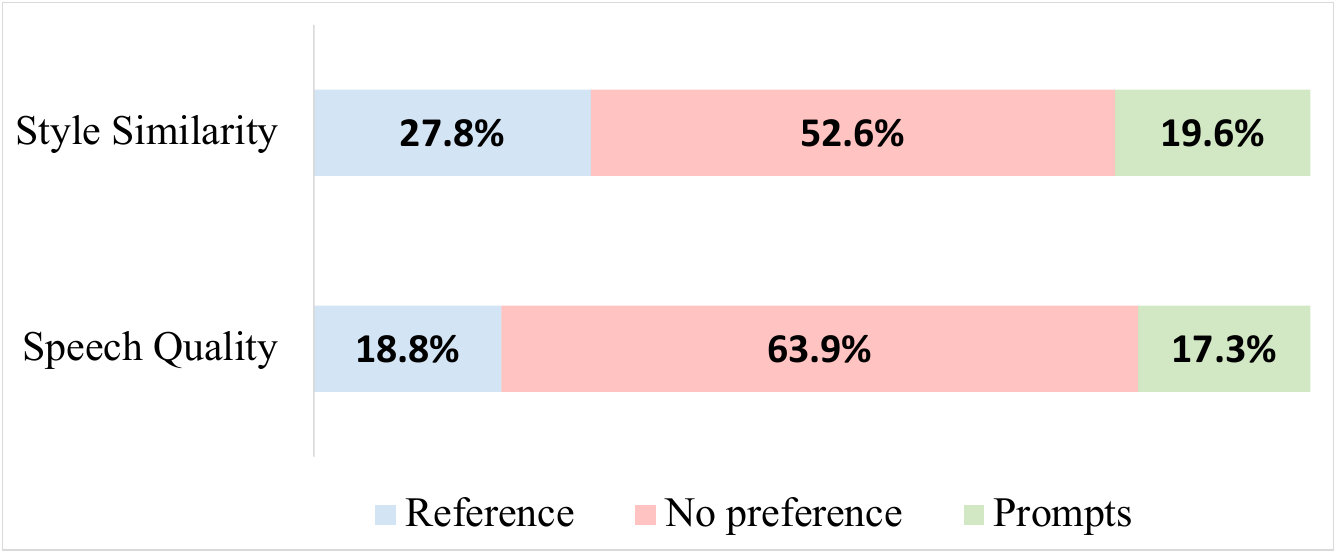}
  \caption{The results of the ABX preference test between reference speech (denoted as Reference) and natural language prompts (denoted as Prompts).}
  \label{fig:abx}
  \vspace{-10pt}
\end{figure}

\vspace{-10pt}
\subsubsection{Style Voice Conversion by Natural Language Prompts}
Given the above observation, PromptVC outperforms the baseline system in style voice conversion using reference speech. Consequently, we proceed to conduct ABX preference tests, which aim to compare the performance of style voice conversion when utilizing reference speech versus employing natural language prompts within our proposed system. 
The first ABX test is utilized to evaluate the degree of style relevance between the generated speech and its corresponding natural language style prompt. Participants are asked to provide a 3-point score ranging from -1 to 1 to choose the speech samples that sound closer to the natural language prompts in terms of style expression. A score of 0 indicates no preference. It is important to note that reference speech serves as the upper bound for the performance of style voice conversion, given that reference speech inherently holds more immediate and direct style information compared to natural language prompts. 

ABX preference test results are shown in Figure~\ref{fig:abx}. In the style similarity test, the results show that 52.6\% of the responses indicated "No preference" and 19.6\% of the responses favored the "Prompts." These results show that PromptVC effectively accomplishes the objective of style control guided by natural language prompts, achieving precise control over style expression. Additionally, we further evaluate the proximity of speech quality between the generated speech and the ground truth sample. Participants must choose the generated speech samples that exhibit a closer alignment to the ground truth sample in terms of speech quality. The results show that the percentage of Reference and Prompts preferences are 18.8\% and 17.3\%, respectively. This suggests that style voice conversion guided by natural language prompts also leads to high speech quality.

\vspace{-10pt}
\subsubsection{Ablation Study}
Since style disentanglement and prosody modeling play an important role in style voice conversion, we conduct ablation studies by replacing the linguistic units from our proposed to the conventionally used PPG, as well as by removing the prosody encoder.  From the results in Table~\ref{fig:abx}, we can draw the following conclusions: (1) In scenarios where the prosody encoder is omitted, the SMOS results are lower than those achieved by our proposed system. This highlights the contribution of the prosody encoder in improving the style similarity and naturalness of the converted speech. (2) A significant decline in SMOS, RMSE and Corr values is observed when the linguistic representation switches from our proposed linguistic units to PPG. This suggests that our proposed method yields superior disentanglement performance and effectively reduces residual style information in the linguistic content. Retained original style information within the linguistic content leads to diminished style similarity in the converted speech. The results of ablation studies further demonstrate the effectiveness of each component within PromptVC.

\vspace{-10pt}
\begin{table}[ht]
\caption{Results for ablation study. "w/o prosody" stands for removing the prosody encoder and "variants with PPG" represents replacing our proposed linguistic representation with PPG conventionally.}
\label{fig:ablation}
\renewcommand\arraystretch{1.2}
\resizebox{1.0\linewidth}{!}{
\begin{tabular}{lcccc}
\hline
\multicolumn{1}{c}{Model}    & RMSE ($\downarrow$) & Corr ($\uparrow$)& MOS ($\uparrow$) & SMOS ($\uparrow$) \\ \hline
Proposed & \textbf{11.06} & \textbf{0.86} & \textbf{3.89$\pm0.05$}      & \textbf{3.87$\pm0.06$} \\
w/o prosody & 11.33 & 0.83 & 3.87$\pm0.06$                    & 3.71$\pm0.07$ \\
variants with PPG & 13.41 & 0.78 & 3.76$\pm0.05$                    & 3.62$\pm0.07$ \\ \hline
\end{tabular}
}
\end{table}

\vspace{-10pt}
\section{Conclusions}
In this study, to explore style voice conversion with a natural language prompt, we propose PromptVC, a latent diffusion model designed to generate the style vector and convert speech consistent with the natural language prompt in style. Compared with previous works in style voice conversion, our proposed approach controls the generation of speech in a more user-friendly way. Furthermore, we propose a novel linguistic content extraction method that leverages HuBERT to extract linguistic content. The extracted token is then deduplicated and re-predicted the duration by a differentiable duration predictor, which can change the duration of the linguistic content on different styles. Experimental results demonstrate that our proposed approach can generate speech with precise style control and high speech quality guided by natural language prompts.

\bibliographystyle{IEEEbib}
\bibliography{strings,refs}

\end{document}